\documentclass{aa} 
\usepackage{graphicx,epsfig,amssymb,amsmath,layout,verbatim,rotating,calc,mathrs
fs,natbib,textcomp}
\usepackage{graphics}
\usepackage{color}
\usepackage{rotate}
\usepackage{float}   
\usepackage{ulem}
\usepackage{verbatim}
\usepackage{txfonts}
\newcommand{\ksk}{km~s$^{-1}$~kpc$^{-1}$ }                                      
 
\newcommand{\dgr}{$^{\circ}$}

\begin{document}

   \title{Building CX peanut-shaped disk galaxy profiles}

   \subtitle{The relative importance of the 3D families of periodic
orbits bifurcating at the vertical 2:1 resonance}

   \author{P.A.~Patsis
          \inst{1}
          \and
          M.~Harsoula\inst{1}%\fnmsep\thanks{Just to show the usage
%          of the elements in the author field}
          }

   \institute{Research Center for Astronomy, Academy of Athens, Soranou
Efessiou 4, 115 27 Athens, Greece\\
              \email{patsis@academyofathens.gr}\\
              \email{mharsoul@academyofathens.gr}      
%         \and 
%             University of Alexandria, Department of Geography, ...\\
%             \email{mharsoul@acadmyofathens.gr}
%             \thanks{The university of heaven temporarily does not
%                     accept e-mails}
             }

   \date{ }

% \abstract{}{}{}{}{} 
% 5 {} token are mandatory
 
  \abstract
  % context heading (optional)
  % {} leave it empty if necessary  
   {We present and discuss the orbital content of a rather unusual  rotating
barred galaxy model, in which the three-dimensional (3D) family, bifurcating
from x1 at the 2:1 vertical resonance with the known ``frown-smile'' side-on
morphology, is unstable.}
  % aims heading (mandatory)
   {Our goal is to study the differences that occur in the phase space structure
at the vertical 2:1 resonance region in this case, with respect to the known,
well studied, standard case, in which the families with the
frown-smile profiles are stable and support an X-shaped morphology.}
  % methods heading (mandatory)
   {The potential used in the study originates in a frozen snapshot of an
$N$-body simulation in which a fast bar has evolved. We follow the evolution of
the vertical stability of the central family of periodic orbits as a function of
the energy (Jacobi constant) and we investigate the phase space content by
means of spaces of section.}
  % results heading (mandatory)
   {The two bifurcating families at the vertical 2:1 resonance region of the new
model change their stability with respect to that of most studied analytic
potentials.  The structure in the side-on view that is directly supported by the
trapping of quasi-periodic orbits around 3D stable periodic orbits has now an
infinity symbol (i.e. $\infty$-type) profile. However, the available sticky
orbits can reinforce other types of side-on morphologies as well.}
  % conclusions heading (optional), leave it empty if necessary 
   {In the new model, the dynamical mechanism of trapping quasi-periodic orbits
around the 3D stable periodic orbits that build the peanut, supports
the $\infty$-type profile. The same mechanism in the standard case
supports the X shape with the frown-smile orbits. Nevertheless, in both cases
(i.e. in the new and in the standard model) a combination of 3D quasi-periodic
orbits around the stable x1 family with sticky orbits can support a profile
reminiscent of the shape of the orbits of the 3D unstable family
existing in each model.}
   \maketitle
%
%-------------------------------------------------------------------

\section{Introduction}
Peanut-shaped bulges in $N$-body simulations have been correlated with the
presence of Inner Lindblad Resonances (ILR) \citep{com90} and are considered to
be part of the bars viewed edge-on \citep{am02}. Many orbital models that have
been developed in order to associate the presence of the boxy bulges in galaxies
with the presence of orbital families have shown that the observed structures
can be built by means of the families introduced in the vertical 2:1 resonance
(hereafter vILR) of rotating barred potentials \citep[see
e.g.][]{pf84,pffr91,petal02}.

Briefly, peanut-building families are introduced in the system at the two nearby
vILRs, where the planar, central family, x1, experiences a double
stability transition. Namely, as the Jacobi constant increases, the initially
stable x1 family becomes simple unstable \citep{cm85} and then returns
back to stable, that is, we have a $S\rightarrow U\rightarrow S$ scheme
\citep[see][]{setal02}. At the $S\rightarrow U$ transition we have a
three-dimensional (3D) stable family bifurcating from x1, while at the
$U\rightarrow S$ transition that follows, a 3D unstable family is introduced in
the system. In conclusion, the presence of a pair of vILRs gives rise to two 3D
families of periodic orbits in the system; one stable, and the other unstable.
%-------------------------------------------------------------------------------
\begin{figure}[ht]
\begin{center}
\resizebox{50mm}{!}{\includegraphics[angle=0]{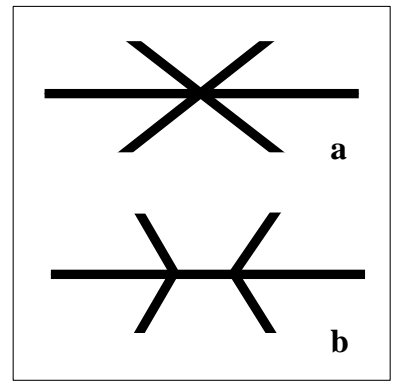}}
\end{center}
\caption{A sketch of the two types of X features appearing in edge-on views of
disk galaxies: (a) The CX type, in which the wings of the X cross the centre of
the system, and (b) the OX profile, in which the wings of X avoid the centre.}
\label{schema} 
\end{figure}  
%-------------------------------------------------------------------------------
%-------------------------------------------------------------------------------
\begin{figure*}
\begin{center}
\resizebox{140mm}{!}{\includegraphics[angle=0]{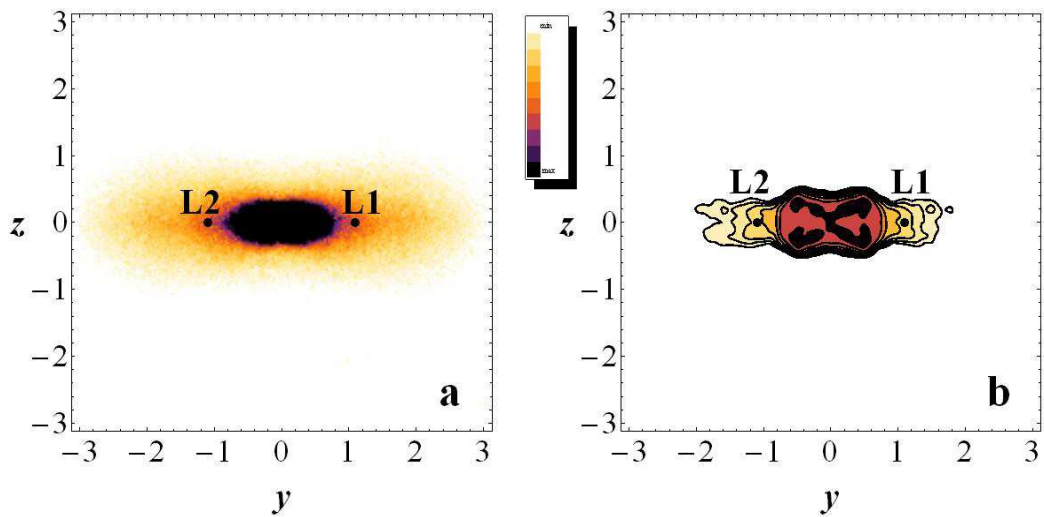}}
\end{center}
\caption{(a) The side-on profile of the $N$-body model. (b) The side on profile
consisting of orbits belonging to the 2:1 class in the corresponding frozen
potential \citep{ch13}. The plotted isodensities in (b) emphasise the CX
character of the profile. L1 and L2 indicate the location of the unstable
Lagrangian points. Darker colour corresponds to more dense regions.}
\label{profil} 
\end{figure*}  
%-------------------------------------------------------------------------------

In the analytic model of the Ferrers bar \citep{pf84,petal02}, in the double
Miyamoto triaxial potential studied by \citet{kpc13} and in the model of
\citet{pffr91} the family of periodic orbits, which is introduced at the
$S\rightarrow U$ transition as stable can be vaguely described as having orbits
resembling `frowns' and `smiles' when viewed side-on (we consider both
branches of the bifurcating family, which are symmetric with respect to the
equatorial plane). In \citet{pffr91} this family changes its stability
at larger energies. An exception to the models in which the  frown-smile orbits
are mainly stable is the model by \citet{muho}. However, even in this case the
family is introduced as stable and becomes unstable at a nearby energy, beyond
but close to the bifurcating point. This family is called BAN by
\citet{pffr91} and x1v1 by \citet{setal02} (its symmetric one being called
x1v1$^\prime$). The combination of the two symmetric with respect to the
equatorial $z=0$ plane side-on projections of x1v1 and x1v1$^\prime$ will give a
`` \raisebox{-.6ex}{{\huge\rotatebox[origin=c]{90}{$\between$}}} '' shape. On
the other hand, the orbits of the 3D family that is introduced in all these
models as unstable (x1v2 in \citet{setal02} and ABAN in \citet{pffr91})
appear in the corresponding projection as `` \raisebox{-.4ex}{{\huge$\infty$}}
''-shaped. \citep[Here we follow the][notation]{setal02}.
 
\citet{betal06} classify the boxy, peanut-shaped, X-type bulges of edge-on disk
galaxies in two morphological classes. They are either CX (centred) or OX
(off-centred), depending on whether or not the wings of the X feature cross the
centre of the galaxy (CX) or not (OX). A sketch describing the two cases of X is
given in Fig.~\ref{schema}. The disk is represented by a horizontal line, since
we deal  with edge-on views. In (a) the wings of X cross the centre of the disk
and thus the sketch describes a CX profile, while in (b) they do not and the
profile is an OX one.

Considering that the standard mechanism that builds the observed profiles is the
trapping of quasi-periodic orbits in the neighbourhood of stable periodic ones,
it is clear that in the studied cases the building of the OX profiles is
favoured in the orbital models, since individual, or a number of, stable x1v1
and x1v1$^\prime$ periodic orbits support this morphology \citep{petal02}.
However, recently \citet{pk14a} emphasised the role of sticky chaotic orbits in
building the peanuts and proposed a possible origin for the CX-type profiles.
According to that study, such profiles can be built as a combination of 3D
quasi-periodic orbits on tori around the planar x1 family together with sticky
chaotic orbits emerging from the x1v2 neighbourhood. The latter orbits are led
by the unstable manifolds of the unstable x1v2  periodic orbits in the region of
the x1v1 tori that flank the x1 island of stability in appropriate $(z,p_z)$
projections of the surfaces of section, if $z$ is the axis of rotation of the
system \citep[see Fig. 14 in][]{pk14a}. These sticky chaotic orbits have
during long time intervals hybrid morphologies between x1v1 and x1v2, in which,
in most studied cases, the morphology of the unstable periodic orbit prevails;
in this case x1v2. Similar portraits of the phase space are encountered in most
orbital models that have been studied. 
 
In the majority of the papers where the dynamics at the vILR region is
discussed, the 3D bar is represented by a triaxial Ferrers potential \citep{fer}
and the stability of the x1v1 and x1v2 families is as described above. On the
other hand, in \citet{ch13}, the order, and thus the stability, of the x1v1 and
x1v2 families is reversed. \citet{ch13} studied the chaotic diffusion in an
$N$-body model that develops a fast rotating bar. They examined the orbital
dynamics in a frozen snapshot of this model in order to show how chaotic orbits,
slowly diffusing from the bar region to the region beyond corotation, can
support the spiral structure. A peculiarity of the frozen snapshot's potential
was that the stable family was x1v2 and the unstable one was x1v1. Thus we say
that their stability has been reversed, compared with the stability of the
corresponding families in the analytic potentials.

In the present paper we examine the consequences that the prevalence of stable
x1v2 ($\infty$-shaped) periodic orbits in galactic bars could have for their
edge-on morphologies. For this purpose we compare the structure of phase space
at the vILR of the frozen $N$-body snapshot in \citet{ch13} with that of a
standard Ferrers bar \citep[][etc.]{pf84, petal02}. As a typical case, we
consider, the one described in \citet{pk14a}. The goal of this study is to trace
the differences in the phase space in the two cases and to find the building
blocks each case offers for building peanut-shaped bulges. In
Sect.~\ref{sec:mod} we briefly describe the model, in Sect.~\ref{sec:phsp} we
present the structure of the phase space of the model at the vILR, in
Sect.~\ref{sec:comp} we compare it with previous models, in
Sect.~\ref{sec:vreson} we compare the variation of the vertical frequencies in
the model by \citet{ch13} with that of a standard Ferrers bar model, and
in Sect.~\ref{sec:concl} we discuss and summarise our conclusions.

\section{A brief description of the model}
\label{sec:mod}
We first briefly describe the model with the unusual stability of the 3D
bifurcating families at the vILR. The $N$-body model from which the potential in
\citet{ch13} originates is described in detail in \citet{hk09}. The code is a
smooth potential field scheme along the line of the \citet{app90} numerical
algorithm. The model develops in about 0.8~Gyr a bar that lasts almost for 1
Hubble time. The snapshot we consider is taken after 55 half-mass crossing times
of the system, which corresponds to about 2.5~Gyr. The $V(x, y, z)$ 3D
'frozen' potential is given by the code as an expansion of a bi-orthogonal
basis set \citep{ch13}. In our analysis we work in the co-rotating with the bar
frame of reference. By using a `frozen' potential, our model is an autonomous
Hamiltonian system, the Hamiltonian of which can be expressed as
%-------------------------------------------------------------------------------
\begin{figure}
\begin{center}
\resizebox{86mm}{!}{\includegraphics[angle=0]{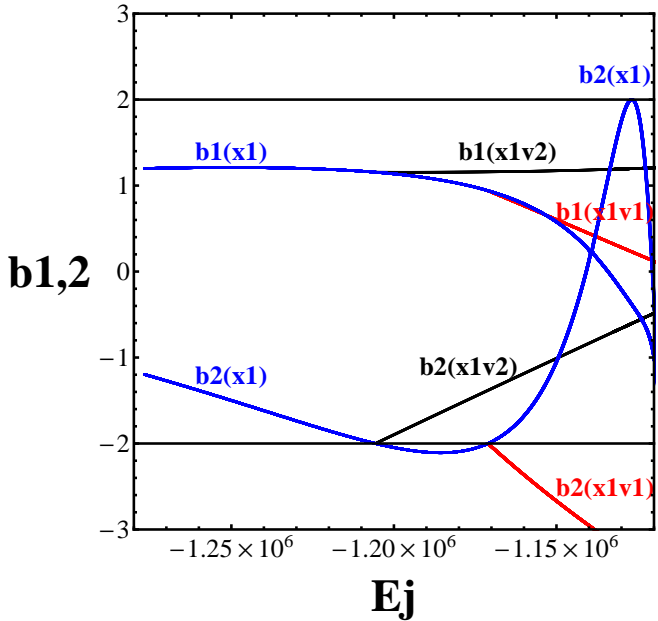}}
\end{center}
\caption{The evolution of the stability indices of the
main families at the vILR
region. From the mother family x1, bifurcate first the x1v2 as stable and
second
the x1v1 as unstable.}
\label{b1b2} 
\end{figure}  
%-------------------------------------------------------------------------------
\begin{equation}
 H = \frac{1}{2}(\dot{x}^2 + \dot{y}^2 + \dot{z}^2 ) + V(x, y, z) -
\frac{1}{2}\Omega_p^2(x^2 + y^2) = E_J,
\end{equation} 
where $(x,y,z)$ are the coordinates in a Cartesian frame of reference rotating
clockwise around the z-axis with angular velocity $\Omega_p$. We have adopted
the clockwise rotation in order to be in agreement with the $N$-body model of
\citet{hk09}. In that simulation, the developed barred-spiral structure rotates
in such a way that the spiral arms are trailing. In the potential $V(x,y,z)$ of
the chosen snapshot the bar is approximately aligned with the y-axis. $E_J$ is
the numerical value of the Jacobi constant, hereafter called the energy and dots
denote time derivatives. The time unit is taken equal to one half-mass crossing
time, while the length unit is taken equal to the half-mass radius
\citep{tetal08}. The adopted value of the pattern speed used in this study
corresponds to $\Omega_p \approx$25~\ksk. In all orbital calculations we have
used a Runge-Kutta seventh order scheme.

The side-on profile of the snapshot of the $N$-body model is given in
Fig.~\ref{profil}a. Heavy dots indicate the location of the unstable Lagrangian
points $L_1$ and $L_2$ along the major axis of the bar (y-axis). Frequency
analysis in the frozen potential \citep{ch13} has shown that this profile is
composed mainly of orbits with frequencies 2:1 and 3:1 on the plane of rotation.
Isolating the particles that follow orbits belonging only to the 2:1 class, that
is, particles on orbits with frequency 2:1 on the equatorial plane, we construct
the profile given in Fig.~\ref{profil}b. The overplotted isodensities clearly
indicate that these orbits build a CX-type profile. Since we have an analytic
potential for this snapshot we can investigate the dynamical mechanisms that
support this structure. 

\section{The vILR phase space structure}
\label{sec:phsp}
The interconnections of the families of periodic orbits in the vILR region can
be followed by means of the stability diagram that gives the evolution of the
stability of the families of periodic orbits as the energy varies \citep{cm85}.
%-------------------------------------------------------------------------------
\begin{figure}[ht]
\begin{center}
\resizebox{85mm}{!}{\includegraphics[angle=0]{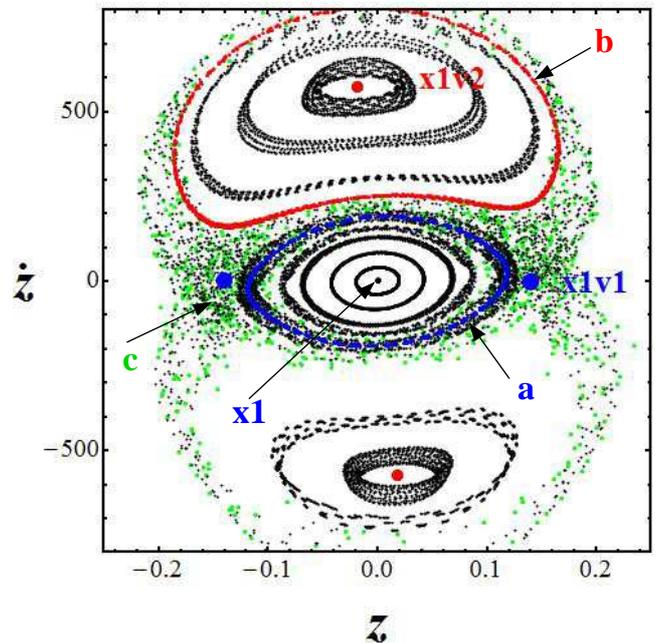}}
\end{center}
\caption{The $(z,\dot{z})$ projection of the space of section for the main
orbits that participate in building the side-on profile of the model, for
$E_J=-1.15 \times 10^6$. Heavy dots indicate the location of x1v1 and x1v2,
while the x1 periodic orbit is projected at $(z,\dot{z})\approx(0,0)$. Arrows
point to the initial conditions of the orbits given in Fig.~\ref{orbs}.}
\label{zpz} 
\end{figure}  
%-------------------------------------------------------------------------------
This is given in Fig.~\ref{b1b2}. Two stability indices, b1 and b2, characterise
the stability of a family with respect to radial and vertical perturbations,
respectively, at a given $E_J$ \citep{br}.

Figure~\ref{b1b2} describes essentially the unusual property of our model that
the x1v2 family is introduced in the system in a smaller $E_J$ than x1v1 as
stable. It is bifurcated at $E_J\approx -1.205 \times 10^6$ from x1 at the
$S\rightarrow U$ transition. The family x1v1 is introduced in the system at the 
%-------------------------------------------------------------------------------
\begin{figure*}[ht]
\begin{center}
\resizebox{180mm}{!}{\includegraphics[angle=0]{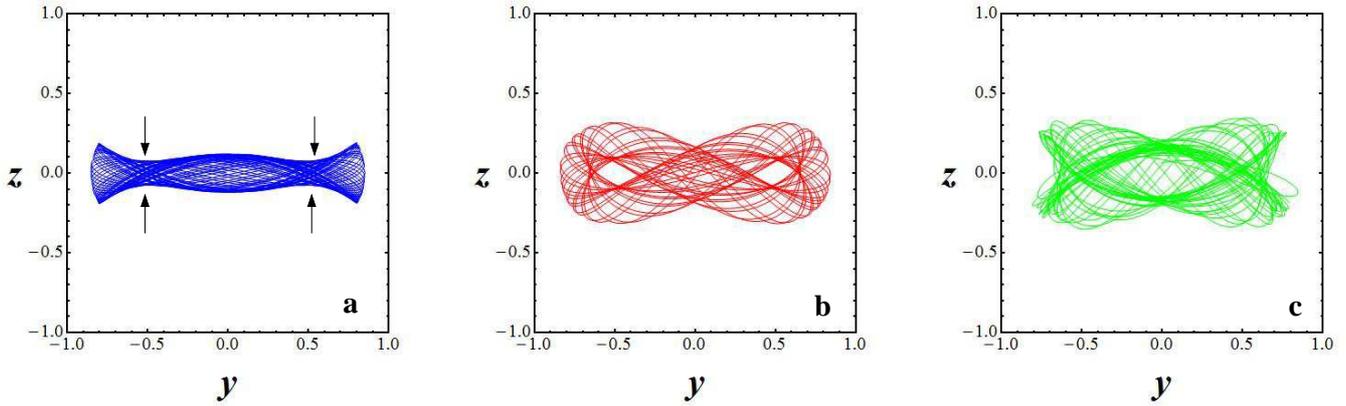}}
\end{center}
\caption{(a) The 3D quasi-periodic orbit around x1 indicated with ``a'' in
Fig.~\ref{zpz}. (b) The quasi-periodic orbit around x1v2 indicated with ``b'' in
Fig.~\ref{zpz}. (c) The sticky chaotic orbit corresponding to the scattered
green consequents in the chaotic zone of the same Figure. All orbits have been
integrated for time giving 30 consequents on the $y=0$ space of section with
$\dot{y}< 0$.}
\label{orbs} 
\end{figure*}
%-------------------------------------------------------------------------------
slightly larger energy $E_J\approx -1.17 \times 10^6$ at the $U\rightarrow S$
transition. The Lagrangian $L_1$ and $L_2$ points are located at $E_J\approx
-1.12 \times 10^6$, unusually close to the vertical 2:1 resonance region. This
is also a notable feature of the model. Both families reach corotation without
changing their stability.

In order to describe the structure of the phase space in the presence of the
three main families that determine the overall morphology of the thick bar, we
chose the energy $E_J=-1.15 \times 10^6$, just beyond the one at which the
second family, x1v1, has been introduced in the system. We consider the surface
of section $y=0$, that is, our consequents are the intersections of the orbits
with the $y=0$ plane with $\dot{y}\leq0$ (chosen to be like this due to the
clockwise rotation) in the six-dimensional phase space. Because of the energy
conservation our space of section is reduced to the four-dimensional space
$(x,z,\dot{x},\dot{z})$. We remind that in symmetric, analytic potentials, the
initial conditions of the x1v2 orbits are characterised by $z_0=0$, $\dot{z}\neq
0$, while those of x1v1 by $z_0\neq 0$, $\dot{z}=0$. Both families have of
course also a non-zero initial position on the equatorial plane, that is, $x\neq
0$ in this case. We note however, that since the potential we study here
originates in an $N$-body model snapshot, the x1 periodic orbits do not start on
the x-axis with $\dot{x}=0$, and the bifurcating families are not perfectly
symmetric with respect to the equatorial plane. Thus the coordinates that would
be zero in the initial conditions of the corresponding families in a symmetric
potential are small in the present model, but in general non-zero. For our
discussion it is convenient to use  $(z,\dot{z})$ projections to describe the
results. We chose this particular projection and the specific orbits in order to
facilitate the description of our results. Inclusion of more orbits existing in
this energy and visualisation of the 4D space of section with other methods as
in \citet{pz94} and \citet{kpc13} would not clarify the structure of phase-space
in the present case.

The $(z,\dot{z})$ projection of the 4D space of section for $E_J=-1.15 \times
10^6$ is given in Fig.~\ref{zpz}. At this energy we have the stable x1 and x1v2
periodic orbits and the unstable x1v1 one. The location of x1 is
$(z,\dot{z})\approx(0,0)$. The elliptical `thick' curves around it are the
projections of tori that surround it \citep{kp11}. More precisely they
are the tori of the quasi-periodic orbits, which we find by perturbing
successively, by increasing $\Delta z$, the x1 initial conditions. Because of
the asymmetry of the potential, the projections of the tori in the $(z,\dot{z})$
plane will have in general a certain thickness. The orbits on these tori in the
configuration space have side-on projections like the one in Fig.~\ref{orbs}a.
This particular orbit corresponds to the torus coloured blue in Fig.~\ref{zpz},
indicated with an arrow labelled with ``a''. All orbits similar to the one in
Fig.~\ref{orbs}a are characterised at $(y,z)\approx(0,0)$ by a local maximum in
$|z|$, while there are two local minima on its sides (indicated with arrows in
Fig.~\ref{orbs}a). With increasing $|z|$ along the $\dot{z}=0$ axis in
Fig.~\ref{zpz} these two local minima in the successive quasi-periodic orbits
tend to reach $z=0$. This happens when we consider the orbit with the x1 initial
conditions, but with $|z|\approx 0.14$ instead of 0. In other words the two
local minima of the quasi-periodic orbits tend to $z=0$ as we approach the
initial conditions of the unstable periodic orbits x1v1 and x1v1$^{\prime}$
moving along the $\dot{z}=0$ axis in Fig.~\ref{zpz}.

Above and below the region occupied by the x1 tori in Fig.~\ref{zpz} we observe
two more regions with tori projected in the $(z,\dot{z})$ plane. They belong to
the quasi-periodic orbits around the two members of the stable x1v2 periodic
orbit that have $\dot{z} \approx \pm 573.64$  in our velocity units,
respectively, and are located close to the $z=0$ axis. The side-on projections
of the quasi-periodic orbits on the x1v2 tori remind us of the $\infty$-type
morphology of the x1v2 periodic orbits. Such orbits are characterised at
$(y,z)\approx(0,0)$ by a local {minimum}  in $|z|$ this time. A typical example
is given in Fig.~\ref{orbs}b, which corresponds to the red orbit indicated
with ``b'' in Fig.~\ref{zpz}.

The $\infty$-type profile is not the only morphology that can be supported
in the side-on view of the model. The sticky chaotic orbits with initial
conditions in the neighbourhood of the unstable x1v1 and x1v1$^{\prime}$
periodic orbits also provide building blocks for long-lasting structures. In
Fig.~\ref{zpz} the consequents that belong to such orbits are roughly projected
around the x1 and x1v2 invariant tori building a chaotic layer. The green
coloured consequents in this region belong to a typical sticky orbit depicted in
Fig.~\ref{orbs}c. Its initial condition in the $(z,\dot{z})$ plane is indicated
with an arrow close to x1v1$^{\prime}$ (Fig.~\ref{zpz}, left side, labelled with
``c''). Evidently this is an orbit with a hybrid morphology between x1v1 and
x1v2. Nevertheless, the x1v1 character prevails. 

\subsection{Composite profiles}
\begin{figure}[ht]
\begin{center}
\resizebox{80mm}{!}{\includegraphics[angle=0]{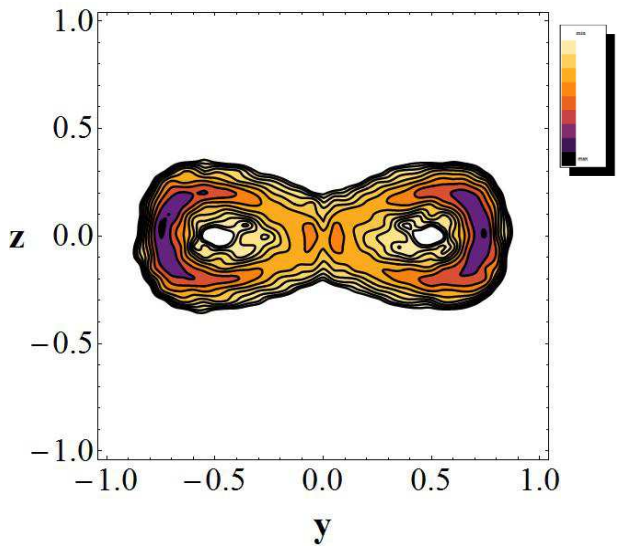}} 
\end{center}
\caption{An image created by overplotting the side-on view of 12
quasi-periodic orbits around x1v2 periodic orbits at 4 different energies. The
isodensity contours indicate the support of a CX-type profile. Density
increases from top to bottom on the colour bar at the right of the
Figure.}
\label{compv2} 
\end{figure}

Besides the individual orbits, we also investigate the morphologies that are
supported by the overlapping of several non-periodic orbits associated with one
family in different energies. A profile composed by the overplotting of 12
quasi-periodic orbits around stable x1v2 periodic orbits is given in
Fig.~\ref{compv2}.
%-------------------------------------------------------------------------------
%-------------------------------------------------------------------------------
We consider three orbits at each of the energies $E_J = -1.16, -1.15, -1.14$ and
$-1.13 \times 10^6$. One of these tori belongs to an inner torus (like the
innermost one in Fig.~\ref{zpz}), another one to a torus just before entering
into the sticky-chaotic zone that surrounds the tori (like the one labelled with
``b'' in Fig.~\ref{zpz}) and in all cases we consider also a third orbit on a
torus between these two. The orbits have been integrated for the time needed to
give 30 consequents in the $y=0$ space of section. In order to make the effect
of the orbital overlapping more discernible, we have converted the
twelve overlapping orbits diagram to an image. The orbits are plotted using a
constant time step. By constructing the image, we consider the intensity of each
pixel to be proportional to the local number density of the points of the orbits
in its region. The plotted curves are isodensities that delineate the supported
structure. As we can observe in Fig.~\ref{compv2} the overlapping of several
quasi-periodic orbits around members of the x1v2 family supports a CX profile.
In this case the role of the orbits on tori close to the periodic orbit is
important, since they support sharper CX-type X features.

\begin{figure}[ht]
\begin{center}
\resizebox{90mm}{!}{\includegraphics[angle=0]{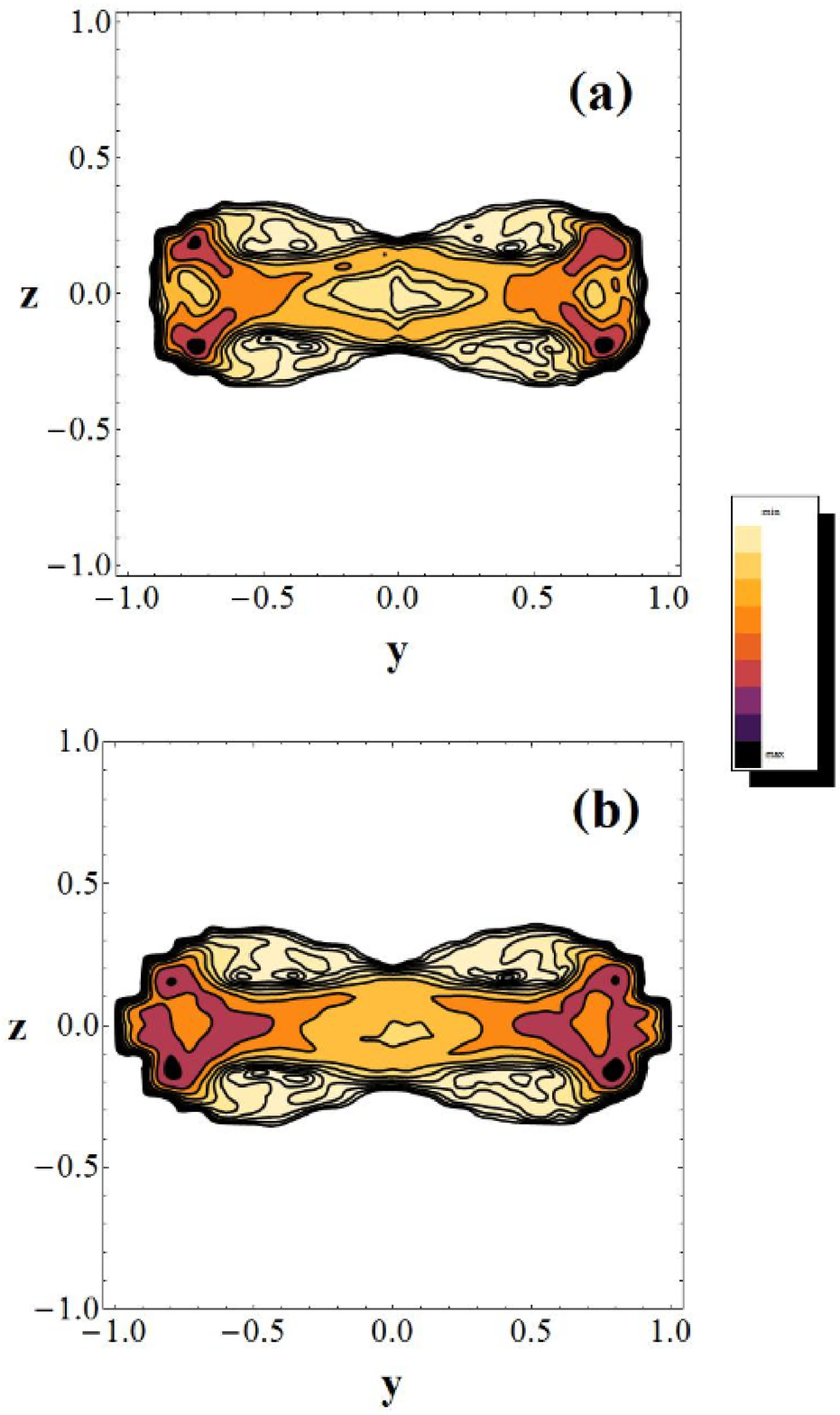}} 
\end{center}
\caption{(a) An image created by overplotting the side-on view of four
sticky-chaotic orbits with initial conditions close to x1v1 periodic orbits at
four different energies. (b) The same image including also four 3D quasi-periodic
orbits around x1. The isodensity contours in both cases indicate the support of
an OX-type profile. Density increases from top to bottom on the
colour bar at the right of the Figure.}
\label{compv1} 
\end{figure}

We can build a similar profile by means of the sticky orbits we find by starting
with initial conditions in the neighbourhood of the unstable periodic orbits
x1v1 and x1v1$^{\prime}$. The topology of the phase-space of the model for
energies $E_J \geq -1.16 \times 10^6$, in the $(z,\dot{z})$ projection, is
similar to the one given in Fig.~\ref{zpz}. Thus, a x1v1-like profile can be
built by such orbits with energies in the range $-1.17 \times 10^6 < E_J < -1.12
\times 10^6$. In this case we can combine four sticky orbits that remain close
to the projected tori in the $(z,\dot{z})$ plane, like the orbit with the green
consequents in Fig.~\ref{zpz}. The result is given in Fig.~\ref{compv1}a. 
%-------------------------------------------------------------------------------
%-------------------------------------------------------------------------------
For constructing this image, we followed the same procedure as in the case of
the image in Fig.~\ref{compv2}. The orbits are taken at $E_J = -1.16, -1.15,
-1.14$ and $-1.13 \times 10^6$. In this case the profile is of OX type, since
the wings of X emerge out of the equatorial plane at $|y|>0.5$. We note that the
wings remain sharp features, despite the fact that we have used sticky chaotic
orbits. The result does not change considerably if we include also 3D
quasi-periodic orbits around x1, that is orbits like the one in
Fig.~\ref{orbs}a, as we can observe in Fig.~\ref{compv1}b. Such orbits are by
themselves of OX type and their inclusion simply continues the same topology to
smaller radii and lower heights.

In summary, the available orbital content for building structures can be either
x1v2-like, mainly due to the quasi-periodic orbits around the x1v2 periodic
orbit, or x1v1-like, because of the existence of sticky chaotic orbits in the
region between and around the invariant tori we encounter in the phase space, at
$E_J$ for which both x1v2 and x1v1 exist. The model offers mainly a
straightforward mechanism for building CX-type side-on profiles by means of
regular orbits such as the one in Fig.~\ref{orbs}b. However, by populating the
bar of the model with 3D quasi-periodic orbits around x1 (Fig.~\ref{orbs}a) and
sticky orbits similar to the one in Fig.~\ref{orbs}c we can support a
frown-smile, OX profile as well.

\section{Comparison with the standard case}
\label{sec:comp}
The structure of phase space we encounter in Fig.~\ref{zpz} can be compared with
the phase space structure in a rotating Ferrers bar model, such as the one in
\citet{pf84}, \citet{petal02} and similar models. In this Section we compare
qualitatively the phase space structure presented in Fig.~\ref{zpz} with that in
the rotating Ferrers bar model studied by \citet{pk14a}.

\begin{figure}[ht]
\begin{center}
\resizebox{90mm}{!}{\includegraphics[angle=0]{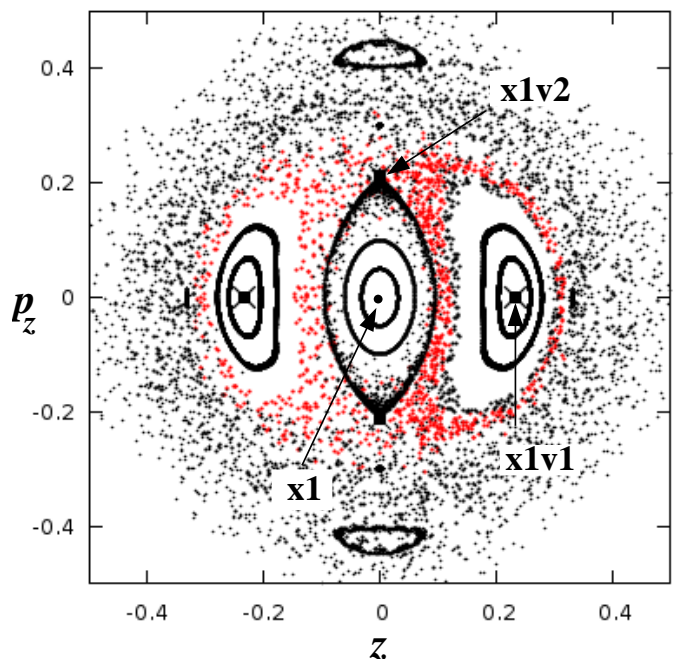}} 
\end{center}
\caption{The $(z,p_z)$ projection of the Ferrers bar model in \citet{pk14a}
corresponding to Fig.~\ref{zpz}. The stability of the 3D families bifurcated
from x1 is now reversed with respect to those of the model from the
$N$-body
simulation. Arrows indicate the locations of the main families of periodic
orbits in this energy.}
\label{f14} 
\end{figure}

An energy for which both 3D families of periodic orbits, which bifurcate from x1
at the vILR exist (i.e. x1v1 and x1v2) is $E_J=-0.41$ in the units of that model
\citep[see][]{pk14a}. The projection of the space of section corresponding to
Fig.~\ref{zpz} is given in Fig.~\ref{f14}. The 4D space of section in this case
is $(x,p_x,z,p_z)$ with equations of motions derived from the Hamiltonian
\begin{equation}
H= \frac{1}{2}(p_{x}^{2} + p_{y}^{2} + p_{z}^{2}) +
    \Phi(x,y,z) - \Omega_{b}(x p_{y} - y p_{x})    .
\end{equation}
The potential $\Phi(x,y,z)$ consists of a Ferrers bar having as axisymmetric
background a Miyamoto disk \citep{mn75} and a Plummer sphere \citep{pl11}. It is
a typical model of a 3D strong bar. Details and numerical values of the
parameters of the components of the potential, scaling of units, and so on, can
be found in \citet{pk14a}. We note that in this case the bar is rotating
counter-clockwise around the z-axis and is aligned with the y axis. Our surface
of section is defined by $y=0$ and we consider the consequents with $p_y>$ 0.
Figures~\ref{zpz} and \ref{f14} have a conspicuous qualitative similarity, when
rotated by 90\dgr. Nevertheless, there is a striking difference, namely that the
x1v1 and x1v2 families of periodic orbits have exchanged their stability
%-------------------------------------------------------------------------------
%-------------------------------------------------------------------------------
and their relative locations. So, the location of the non-periodic orbits in
phase space has changed accordingly as well. 

In the configuration space, the side-on views of the quasi-periodic orbits in
the tori drawn around $(z,p_z)=(0,0)$, that is, of the 3D regular orbits around
x1, are now of $\infty$-type, instead of having the morphology of
Fig.~\ref{orbs}a. By inspection of Fig.~\ref{f14} we can see that successive
projections of the tori of the 3D quasi-periodic orbits around x1, encountered
as we perturb the x1 initial condition in the $p_z$ or $z$ direction, approach
the x1v2 initial condition. Their morphology now resembles that of the x1v2
periodic orbit and can be considered of similar morphology to the orbit in
Fig.~\ref{orbs}b \citep[cf. Fig. 15e in][]{pk14a}. 

The quasi-periodic orbits on the tori around x1v1 and x1v1$^{\prime}$ (on the
sides of the x1 stability island in Fig.~\ref{f14}) support frown and smiles
morphologies. They are not of $\infty-$type as in the case of 3D tori of
Fig.~\ref{zpz}. A set of them in the energy range in which x1v1 exists will
clearly reinforce an OX peanut-shape morphology \citep[cf. Fig. 9a
in][]{petal02}. On the other hand, chaotic orbits sticky to these tori, like the
one plotted with the red consequents in Fig.~\ref{f14}, have a hybrid character
in the configuration space, in which frequently the $\infty$-type morphology
prevails \citep[cf. Fig. 13 in][]{pk14a}. In order to obtain a basic idea of the
structures that are supported by sticky-chaotic and quasi-periodic orbits in the
Ferrers bar model, we may look at Fig.~\ref{orbs}b and Fig.~\ref{orbs}c,
respectively. We note that there is an overall symmetry as regards the orbital
content in the two models, however, with the role of the 3D bifurcated families
in providing quasi-periodic or sticky orbits being reversed.

\section{Variation of the vertical resonances in the two models}
\label{sec:vreson}
A basic difference between the two models is the variation of their vertical
frequencies. This leads to totally different ``relative'' energies at which the
vertical resonances appear. In order to express this in a way to allow
comparison of the two models, we calculate the quantity $E_* = \Big| \dfrac{E_J
- E_0 }{E_{L_{1}} - E_0} \Big |$, where $E_J$ is the energy at the location of a
vertical resonance,  $E_{L_{1}}$ is the energy of the $L_1$ Lagrangian point and
$E_0$ is the energy at the bottom of the effective potential well, at
$(x,y,z)=(0,0,0)$. The relative energy, $E_*$, gives us a measure of how
close to corotation appears a vertical resonance in each model. We chose
$E_{L_{1}}$ to represent the corotation energy, because we want to follow the
behaviour of the two models along the y-axis, that is, along the major axis of
the bar.  

The most accurate way to locate resonances in strongly non-axisymmetric models
like the two models of barred galaxies we compare, is the variation of the
stability indices of their central family of periodic orbits \citep[see
e.g.][]{cg89, pg96, gcobook02}. In stability diagrams of axisymmetric
models, the vertical stability index becomes tangent to the $-2$ axis, exactly
at the energies of the vertical resonances. However, in the full potential,
these energies are considerably displaced from the location of the resonances in
the axisymmetric background. In the presence of barred potentials, at the
vertical resonances, two sets of new families of periodic orbits appear
bifurcating from the bar-supporting ellipses on the plane. An example are the
families bifurcated from x1 in Fig.~\ref{b1b2}, as the vertical stability index
(b$_2$) intersects the $-2$ axis. These regions are characterised by a
successive $S\rightarrow U\rightarrow S$ transition of the vertical stability of
\begin{table*} 
  \begin{center}
    \caption{Comparison of the location of the vertical resonances in the two
models along the major axis of the bar. Each row successively gives the name of
the vertical resonance, the relative energy $E_*$ (see text) 
of each resonance 
for the $N$-body model and the Ferrers bar respectively and finally the maximum
length of the x1 orbits at the resonance (their apocentra), normalized over the
$L_1$ radius,
for the two models.}
    \label{tab:table1}
    \begin{tabular}{c|c|c|c|c}  
      \textbf{Vertical resonances} & \textbf{$E_*$} & \textbf{$E_*$}& 
      \textbf{$y_m/r_{L_{1}}$}& \textbf{$y_m/r_{L_{1}}$}\\
      & \textbf{($N$-body)} & \textbf{(Ferrers)} & \textbf{($N$-body)}  &  
        \textbf{(Ferrers)} \\
      \hline
      2:1 (a) & 0.921& 0.334& 0.694& 0.125\\
      2:1 (b) & 0.952& 0.370& 0.745& 0.146\\
      \hline
      3:1 (a) & 1.003& 0.515& 0.912& 0.254\\ 
      3:1 (b) & -    & 0.558& -    & 0.287\\ 
      \hline
      4:1 (a) & -    & 0.866& -    & 0.597\\ 
      4:1 (b) & -    & 0.895& -    & 0.633\\ 
      \hline
    \end{tabular}
  \end{center}
\end{table*}
the x1 family. These transitions can give us the characteristic energies at
which the system ``feels'' the resonances.

In order to associate these energies with characteristic lengths along an axis,
we need to link them with a suitable family of periodic orbits. Since we study
bars, we have considered the x1 bar-supporting family on the plane and the
apocentra of their orbits. In the symmetric Ferrers bar model, these apocentra,
like the Lagrangian points $L_1$ and $L_2$, are along the major axis of the bar,
while in the model from the $N$-body snapshot, the x1 apocentra are almost on
the major axis.

We have summarised all the information for the two models we compare in
Table~\ref{tab:table1}. In the first column we give the n:1 vertical resonances
at which the bifurcated 3D families are introduced in the system. In the second
and third columns we give the quantity $E_*$, which indicates the energy level
at which a vertical resonance appears. By definition $E_* =1$ at $E_J=E_{L_1}$.
Thus, the larger $E_*$, the closer to corotation we have the resonant $n$:1
family. $E_*$ is given successively for the potential from the $N$-body snapshot
and for the Ferrers bar model. Finally in the fourth and fifth columns we give
the ratios $y_m/r_{L_{1}}$ for the two models, which specify the apocentra of
the x1 orbits at the bifurcating energies in which the $n$:1 resonant 3D
families are introduced. For the $N$-body potential $r_{L_{1}}=1.1$, while
in the Ferrers bar model $r_{L_{1}}=6.56$.

There are conspicuous differences as regards the location of the
vertical resonances in the two models. In the $N$-body potential the two
resonant 3D families associated with the 2:1 vertical resonance appear at energy
levels that are more than 90\% of the effective potential well height,
counting from its minimum to the $L_1$ energy (see $E_*$ column in
Tab.~\ref{tab:table1}). Contrarily, for the corresponding vertical 2:1 families
of the Ferrers bar potential, $E_* \approx 0.35$. This means that the basic
families, which in both models shape the peanut, are introduced in quite different
energy levels. In order to get a better understanding of this difference, we
compare the quantities $y_m/r_{L_{1}}$ for the two models, that is, the apocentra
of the bar-supporting periodic orbits at this energy, normalized over the
distance from the centre of the system to $L_{1}$. In both models the apocentra
and the $L_{1}, L_{2}$ Lagrangian points can be considered to be along the major
axis of the bar. We observe that while in the $N$-body potential the
peanut-supporting 3D periodic orbits start existing along the major axis of
the bar at about 70\% of the $r_{L_{1}}$ radius, in the Ferrers bar model the
corresponding distance from the centre is only 12.5\% of $r_{L_{1}}$. In the
latter case the higher-order vertical resonances 3:1 and 4:1 introduce new 3D
families of periodic orbits away from the end of the bar. Their orbits are more
elongated and remain closer to the equatorial plane, shaping this way a
composite stair-type vertical profile \citep{petal02}. This does not hold in the
case of the $N$-body potential. We observe in Table~\ref{tab:table1} that
even for the 3:1 vertical resonance, $E_*=1.003$, meaning that this resonance
appears at an $E_J$ between the energies of the Lagrangian points ($E_{L_1} <
E_J < E_{L_4}$). At this energy we encounter already orbits in the bar region,
which can cross corotation.

Since the vertical profiles of the bars are determined by the location
of the vertical resonances and by the orbital patterns introduced by them in the
system, it is clear from Table~\ref{tab:table1} that the two cases we compare are
expected to be totally different in that respect. Indeed, the
vertical structure of the $N$-body snapshot is determined almost exclusively
by 3D 2:1 orbits introduced in the system very close to corotation. This leads
to a fast peanut-shaped bar (corotation-to-bar ratio $R_c/R_b \approx 1.1$), in
which the peanut structure is the bar itself, as we can
observe in Fig.~\ref{profil} \citep[cf. also with Fig. 1 in][]{ch13}.
Contrarily, the peanut of the Ferrers bar model is located in the central region
of a fast bar ($R_c/R_b \approx 1.35$), which can be composed also of orbits
trapped around narrow 3D periodic orbits bifurcated at higher-order vertical
$n$:1 resonances $(n>2)$ \citep[see Fig. 12 in][] {pk14b}.

The different vertical profiles of the two models could be the cause of the
differences we encounter in the stability of the families of periodic orbits,
which are introduced at the vertical 2:1 resonance, and consequently of the
differences in the phase-space structure at the region. However, for the goals
of the present study, most important is the fact that despite these differences,
the vertical 2:1 resonance in both cases offers orbital building blocks that
support the peanut and the X structure.

\section{Discussion and Conclusion}
\label{sec:concl}
The case of the model from the frozen $N$-body snapshot we present here
is
rather unusual among the analytic models used so far to study the orbital
dynamics in the vILR region of 3D rotating bars. Its peculiarity consists in the
fact that the stable 3D families of periodic orbits introduced in the system
trap around them quasi-periodic orbits that support side-on CX profiles. On the
other hand, in the majority of the orbital models found in the relevant
literature, the stable 3D families support side-on OX profiles as a
superposition of non-periodic orbits with a frown-smile character. The stability
of one or the other family and the subsequent trapping of quasi-periodic orbits
around them could be the dynamical mechanism that explains the appearance of
either CX or OX profiles in observed galaxies and in $N$-body models. However,
the amount of chaos in the four-dimensional surfaces of section may play the
most important role in both cases. This is indicated both in \citet{kpc13} and
in \citet{pk14a} and is also confirmed in the present study that presents also a
different phase space structure in a vILR region. In all cases, the chaotic zone
occupies a considerable volume in phase space, while it remains close to
invariant tori belonging to x1 and either the stable x1v1 and x1v1$^{\prime}$ or
the set of two x1v2 orbits. Despite the fact that the surface of section has
four dimensions, the presence of these tori close to each other (Figs.~\ref{zpz}
and \ref{f14}) create zones of stickiness that are able to keep the particles in
the region for times of the order of several Gyr and thus bring into the system
orbits that can support structures within this time period\footnote{We note that
in the 3D Ferrers bar model there is one more set of stable tori, as we can
observe in the upper and lower part of Fig.~\ref{f14} for $z=0$, that has its
own sticky zone of influence (consequents in the periphery of the chaotic region
that surrounds the stability islands) and gives a third alternative for a
side-on morphology -- see Patsis \& Katsanikas (2014). However, we do not
discuss this case in the present study.}. Several papers use two-dimensional
frequency analysis in order to classify the observed orbital morphologies
\citep[see e.g.][]{ck07, vhc07, hk09, vsad16, wam16, avsd17}. However, in the
presence of sticky (weakly chaotic) orbits, it is difficult to attribute the
observed structure to a specific 3D resonant family of periodic orbits. Hybrid
morphologies will be present and a detailed investigation of the phase space
structure is needed in order to identify the prevailing contribution by one of
them.

Both CX and OX profiles can be built by orbits introduced in the vILR
resonance of the two models. The existence of vertical resonances is not a
particular property of a model. Vertical resonances exist in 3D
rotating potentials and the vertical 2:1 resonance is a basic one. By
investigating the conditions under which a CX or OX profile will be
formed, we realise that by populating the model with sticky orbits together with
the 3D quasi-periodic orbits trapped in the neighbourhood of x1, one can create
a profile based on the presence of the unstable periodic orbits, instead of the
one expected by the prevalence of the quasi-periodic orbits of the stable 3D
family in each case. Thus, the model of the frozen snapshot of the $N$-body
model of \citet{ch13} can build a CX profile by means of quasi-periodic
orbits around the stable x1v2 periodic orbit and an OX profile using the
quasi-periodic orbits around x1 combined with the sticky chaotic orbits
emanating from the neighbourhood of the unstable x1v1. On the other hand, in the
standard case of a rotating Ferrers bar, like the one in \citet{pk14a}, the
role of quasi-periodic and chaotic orbits associated with the x1v1 and x1v2 is
reversed. However, both models provide the building blocks for each type of
profile. The present study underlines the flexibility that exists 
in dynamical mechanisms, which are based on 3D orbits ``born'' at the vILR of a 
rotating bar, in supporting boxy/peanut-shaped bulges.

In summary:
\begin{itemize}
 \item In the standard case regular orbits support OX profiles and 
sticky chaotic CX ones.
 \item In cases like the potential from the \citet{ch13} model, sticky chaotic 
orbits support OX profiles and regular orbits CX ones. This is valid
also for 
the energies in which the x1v2 family becomes stable, while x1v1 is unstable, 
in any model.
 \item It is the chaoticity of the profiles that decides which dynamical
mechanism will prevail. Definitely, in both cases the vILR region provides
dynamical mechanisms for building X-shaped central regions in barred galaxy
models.
\end{itemize}
 
\begin{acknowledgements}
This work was partly supported by the Research Committee of the Academy of
Athens, project number 200/854. We acknowledge fruitful discussions with  
G.~Contopoulos and L. Athanassoula.
\end{acknowledgements}

% WARNING
%-------------------------------------------------------------------
% Please note that we have included the references to the file aa.dem in
% order to compile it, but we ask you to:
%
% - use BibTeX with the regular commands:
%   \bibliographystyle{aa} % style aa.bst
%   \bibliography{Yourfile} % your references Yourfile.bib

\begin{thebibliography}{}
% 
%   \bibitem[Baker(1966)]{baker} Baker, N. 1966,
%       in Stellar Evolution,
%       ed.\ R. F. Stein,\& A. G. W. Cameron
%       (Plenum, New York) 333
\bibitem[Abbot et al. (2017)]{avsd17} Abbott, C. G., Valluri, M., Shen, J.,
Debattista, V. P., 2017, \mnras, 470, 1526
\bibitem[Allen et al. (1990)]{app90} Allen, A. J., Palmer, P. L., Papaloizou, J.,
1990, \mnras, 242, 576
\bibitem[Athanassoula \& Misiriotis (2002)]{am02} Athanassoula E., Misiriotis A.,
2002, \mnras, 330, 35
\bibitem[Broucke (1969)]{br} Broucke R., 1969, NASA Techn. Rep., 32, 1360
\bibitem[Bureau et al. (2006)]{betal06} Bureau, M., Aronica, G., Athanassoula,
E., Dettmar, R.-J., Bosma, A., Freeman, K. C., 2006, \mnras, 370, 753
\bibitem[Ceverino \& Klypin (2007)]{ck07} Ceverino, D.,  Klypin A., 2007,
\mnras, 379, 1155
\bibitem[Combes et al. (1990)]{com90} Combes, F., Debbasch, F., Friedli,
D., Pfenniger, D., 1990, \aap, 233, 82
\bibitem[Contopoulos (2002)]{gcobook02} Contopoulos, G., ``Order and chaos in 
dynamical astronomy'', Springer, Berlin
\bibitem[Contopoulos \& Grosbol (1989)]{cg89} Contopoulos, G., Grosbol, P., 
1989, A\&ARv, 1, 261
\bibitem[Contopoulos \& Harsoula (2013)]{ch13} Contopoulos, G., Harsoula, M., 2013,
\mnras, 436, 1201
\bibitem[Contopoulos \& Magnenat (1985)]{cm85} Contopoulos, G., Magnenat, P.,
1985, Celest. Mech., 37, 387
\bibitem[Ferrers (1870)]{fer} Ferrers, N. M. 1870, Royal Society of London
Philosophical Transactions Series I, 160, 1
% \bibitem[Harsoula et al. (2011)]{hetal11} Harsoula, M., Kalapotharakos, C.,
% Contopoulos, G., 2011,  \mnras, 411, 1111
\bibitem[Harsoula \& Kalapotharakos (2009)]{hk09} Harsoula, M., Kalapotharakos,
C., 2009, \mnras, 394, 1605
\bibitem[Katsanikas \& Patsis (2011)]{kp11} Katsanikas, M., Patsis, P. A.,
 2011, Int. J. Bif. Chaos, 21, 467
\bibitem[Katsanikas et al. (2013)]{kpc13} Katsanikas, M., Patsis, P. A., Contopoulos,
G., 2013, Int. J. Bif. Chaos, 23, 1330005
\bibitem[Miyamoto \& Nagai (1975)]{mn75} Miyamoto, M., Nagai, R., 1975,
\pasj, 27, 533
\bibitem[Mulder \& Hooimeyer (1984)]{muho} Mulder, W.~A., Hooimeyer, J.~R.~A.,
1984, \aap, 134, 158
\bibitem[Patsis \& Grosbol (1996)]{pg96} Patsis, P.A., Grosbol, P., 1996, \aap, 
315, 371
\bibitem[Patsis \& Katsanikas (2014a)]{pk14a} Patsis, P.A., Katsanikas, M.,
2014, \mnras, 445, 3525
\bibitem[Patsis \& Katsanikas (2014b)]{pk14b} Patsis, P.A., Katsanikas, M.,
2014, \mnras, 445, 3546
\bibitem[Patsis \& Zachilas (1994)]{pz94} Patsis, P.A., Zachilas, L., 1994, Int. J.
Bif. Chaos, 4, 1399
\bibitem[Patsis et al. (2002)]{petal02} Patsis, P.A., Skokos, Ch., Athanassoula,
E., 2002, \mnras, 337, 578
\bibitem[Pfenniger (1984)]{pf84} Pfenniger, D., 1984, \aap, 134, 373
\bibitem[Pfenniger \& Friedli (1991)]{pffr91}Pfenniger, D., Friedli, D., 1991,
\aap, 252, 75 
\bibitem[Plummer (1911)]{pl11} Plummer, H.C., 1911, \mnras, 71, 460
\bibitem[Skokos et al. (2002)]{setal02} Skokos, Ch., Patsis, P.A., Athanassoula,
E., 2002, \mnras,  333, 847
\bibitem[Tsoutsis et al (2008)]{tetal08} Tsoutsis, P., Efthymiopoulos, C.,
Voglis, N., 2008, \mnras, 387, 1264 
\bibitem[Valuri et al. (2016)]{vsad16} Valluri, M., Shen, J., Abbott, C.,
Debattista, V. P., 2016, \apj, 818, 141
\bibitem[Voglis et al. (2007)]{vhc07} Voglis, N., Harsoula, M., Contopoulos,
G., 2007, \mnras, 381, 757
\bibitem[Wang et al. (2016)]{wam16} Wang, Y., Athanassoula, E., Mao, S., 2016,
\mnras, 463, 3499
\end{thebibliography}
%
% - join the .bib files when you upload your source files
%-------------------------------------------------------------------

\end{document}